\documentclass[conference, 10pt, letterpaper]{IEEEtran}
\makeatletter
\def\ps@headings{%
\def\@oddhead{\mbox{}\scriptsize\rightmark \hfil \thepage}%
\def\@evenhead{\scriptsize\thepage \hfil \leftmark\mbox{}}%
\def\@oddfoot{}%
\def\@evenfoot{}}
\makeatother
\IEEEoverridecommandlockouts
\usepackage{cite}
\usepackage{amsmath,amssymb,amsfonts}

\usepackage{enumitem}
\usepackage{graphicx}
\usepackage{textcomp}
\usepackage{xcolor}
\usepackage{hyperref}
\usepackage{comment}
\usepackage{acronym}
\usepackage{algorithm}
\usepackage[noend]{algpseudocode}
\usepackage{cleveref}
\usepackage{soul}
\usepackage{multicol}
\usepackage{array}
\usepackage{tabularx}
\usepackage{pgfplots}
\pgfplotsset{compat=1.18}
\usepackage{filecontents}

\usepackage{cleveref}
\usepackage{cite}
\usepackage{booktabs}

\usepackage{caption} 

\usepackage{multirow}

\usepackage{pgfplotstable}

\usepackage{caption}
\usepackage{adjustbox}
\usepackage{siunitx}
\usetikzlibrary{patterns}

\newtheorem{remark}{Remark}[section]

\newcommand{\group}[1]{\if\relax\detokenize{#1} \mathbb{Z}_{} \else\mathbb{Z}_{#1}\fi}

\newcommand{\extVec}[2]{({#1}_1,\dots,{#1}_{#2})}

\newcommand{\scalProd}[2]{\langle \mathbf{#1}, \mathbf{#2} \rangle}
\newcommand{\scalProdRes}[3]{\sum_{i=1}^{#3} {#1}_{i}{#2}_i}

\newcommand{\Cb}[1]{\textcolor{cb_cyan}{#1}}
\newcommand{\Cr}[1]{\textcolor{cb_caramel}{#1}}

\newcolumntype{P}[1]{>{\centering\arraybackslash}p{#1}}

\definecolor{cb_india}{HTML}{e5af53}
\definecolor{cb_cyan}{HTML}{0173b2}
\definecolor{cb_aquamarine}{HTML}{029e73}
\definecolor{cb_lilac}{HTML}{cc78bc}
\definecolor{cb_caramel}{HTML}{C91F37}
\definecolor{cb_dandelion}{HTML}{ece133}
\definecolor{cb_yellow}{HTML}{ede132}
\definecolor{cb_blue}{HTML}{0072b2}
\definecolor{cb_red}{HTML}{d55f01}

\definecolor{Charcoal Gray}{HTML}{1C1C1C}
\definecolor{Lavender Gray}{HTML}{6E75A4}
\definecolor{Prussian Blue}{HTML}{113285}
\definecolor{Tapestry Blue}{HTML}{0C4842}
\definecolor{Marine Blue}{HTML}{0D5661}
\definecolor{Midnight Blue}{HTML}{0B1013}
\definecolor{Smoke Blue}{HTML}{577C8A}
\definecolor{Marigold}{HTML}{FFB11B}
\definecolor{Raspberry Red}{HTML}{8E354A}
\definecolor{Tomato Red}{HTML}{F75C2F}
\definecolor{Malachite Green}{HTML}{227D51}

\newif\ifFIXESON
\FIXESONtrue 
\ifFIXESON
\newcommand{\fixn}[2]{\fixfootnote{\textbf{#1:} #2}}
\usepackage[draft,inline,nomargin,index]{fixme}
\else 
\newcommand{\fixn}[2]{}
\usepackage[final]{fixme}
\fi
\fxsetup{theme=color,mode=multiuser}
\FXRegisterAuthor{ms}{envms}{\color{Marine Blue}MS}   
\FXRegisterAuthor{gs}{envns}{\color{Smoke Blue}NS}  
\FXRegisterAuthor{at}{envat}{\color{Lavender Gray}AT}   
\FXRegisterAuthor{af}{envaf}{\color{red}AF}   
\FXRegisterAuthor{ft}{envft}{\color{Smoke Blue}GS}  

\fxsetup{layout={footnote}}
\fxusetheme{colorsig}

\acrodef{ddh}[DDH]{Decisional Diffie Hellman}
\acrodef{ddos}[DDoS]{Distributed Denial of Service}
\acrodef{fvg}[FedAVG]{Federated Averaging}
\acrodef{fl}[FL]{Federated Learning}
\acrodef{flad}[FLAD]{adaptive Federated Learning Approach to DDoS attack detection}
\acrodef{fe}[FE]{Functional Encryption}
\acrodef{dl}[DL]{Deep Learning}
\acrodef{he}[HE]{Homomorphic Encryption}
\acrodef{ics}[ICS]{Industrial Control System}
\acrodef{iot}[IoT]{Internet of Things}
\acrodef{ipfe}[IPFE]{Inner Product Functional Encryption}
\acrodef{lwe}[LWE]{Learning With Errors}
\acrodef{mife}[MIFE]{Multi-Input Functional Encryption}
\acrodef{ml}[ML]{Machine Learning}
\acrodef{mlp}[MLP]{Multi-Layer Perceptron}
\acrodef{mpc}[MPC]{Multi Party Computation}
\acrodef{nids}[NIDS]{Network Intrusion Detection System}
\acrodef{tpa}[TPA]{Third Party Authority}

\begin{document}

\title{Adaptive Federated Learning with Functional Encryption: A Comparison of Classical and Quantum-safe Options}

\author{
   Enrico Sorbera$^{\alpha\beta}$,
   Federica Zanetti$^\alpha$,
   Giacomo Brandi$^\alpha$,\\
   Alessandro Tomasi$^\alpha$,
   Roberto Doriguzzi-Corin$^\alpha$,
   Silvio Ranise$^{\alpha\beta}$\\
    
   \small{$^\alpha$Cybersecurity Center, Fondazione Bruno Kessler, Italy, $^\beta$Universit\`a degli Studi di Trento, Italy}
    
}


\maketitle
\begin{abstract}
\ac{fl} is a collaborative method for training aggregate machine learning models while preserving the confidentiality of individual participant training data. Nevertheless, \ac{fl} is vulnerable to reconstruction attacks exploiting shared parameters to reveal private training data. Cryptographic techniques applied to mitigate this threat either incur high computational cost, require sharing private keys, or add extra communication rounds among participants.

In this paper we apply \ac{mife} to a recent \ac{fl} implementation for training Deep Learning-based network intrusion detection systems. We assess both classical and post-quantum solutions in terms of memory and computational overhead. We find that post-quantum algorithms are more computationally efficient in selective security settings but require considerable memory in adaptive security settings.
\end{abstract}

\section{Introduction}
\acf{fl} is a paradigm for training \ac{ml} models with distributed datasets. \ac{fl} allows a federation of multiple parties, called \textit{clients}, to train a common model without requiring them to share the training data.
Although \ac{fl} aims at preserving the confidentiality of client training data, clients are exposed to reconstruction attacks carried out by a malicious aggregator (or \textit{server} in the \ac{fl} terminology). Such attacks can infer details of the clients' original training data \cite{10.5555/3495724.3497145} by exploiting the knowledge of the global model's architecture and information shared by clients during the training process. This includes global model parameters (weights and biases)
and gradients \cite{zhu2019deep}, number of training samples, etc. 

During a reconstruction attack, the malicious server observes the gradient or weight updates sent by the clients and uses optimisation techniques or machine learning models to infer the data that produced them.
The result could be a close approximation of the original training data.

In the current scientific literature this issue has been tackled by adding a cryptographic layer over \ac{fl} with the use of \textit{\ac{fe}}\cite{xu_hybridalpha_2019}, \textit{\ac{mpc}} \cite{mpc} or \textit{\ac{he}}\cite{ZHU202511, fedmlhe}. However, \ac{mpc} has been shown to have high computation and communication costs and, unlike \ac{he}, it requires direct interaction between the \ac{fl} clients \cite{sok_mansouri_2023}. On the other hand, due to its structure, \ac{he} requires all clients to have the same private key. This limitation can be addressed at the cost of introducing additional communication rounds \cite{fedmlhe}. 

In this paper, we address the issue of reconstruction attacks in the cybersecurity domain by introducing \ac{fe} techniques into \ac{fl}, specifically by integrating Multi-Input (Inner Product) Functional Encryption (\ac{mife}) within the \ac{fl} process.
Unlike \ac{he}, \ac{mife} naturally supports different private keys for each client. Our choice of \ac{mife} is motivated by the fact that it integrates seamlessly within the \ac{fl} while requiring minimal communication overhead. 

\ac{mife} enables the \ac{fl} server to execute partial computations on encrypted data across multiple inputs without revealing the underlying plaintext. 
The multiple inputs are the clients' parameters (weights and biases of a \ac{dl} model) and other metadata sent by the clients to the server and used for the management of the \ac{fl} process (e.g., weighted aggregation of clients' updates or client selection). As the server only operates on encrypted data, the clients' original training data remain protected from reconstruction attacks. 


In this work, we focus on two types of algorithms: one whose security relies on the \textit{\acf{ddh}} problem, and another based on the \textit{\acf{lwe}} problem. This comparison allows us to evaluate a classical cryptographic algorithm -- commonly used in the \ac{fl} context -- against a post-quantum one, for which, to the best of our knowledge, no benchmarks currently exist. Additionally, we examine two variants of each algorithm, corresponding to different security notions: selective security and adaptive security, defined in Section \ref{sec:background}. 


We tested classical and post-quantum algorithms in the context of network intrusion detection, where reconstruction attacks can enable the server to obtain sensitive information from the network traffic used to train a \ac{nids} \cite{chen2023feddef}, such as URLs, IP addresses, communication protocols, or even payload fragments. To this aim, we used a recent \ac{fl} solution for training \ac{dl}-based \acp{nids} called \textit{\ac{flad}} \cite{flad,flad_perf}. Compared to the traditional \ac{fl} algorithm, \ac{flad} introduces a mechanism in which the clients share the accuracy score of the global model, as measured on their local validation sets, with the server. 


Experimental results show that, although adding a security layer significantly impacts the memory and time performance of \ac{flad}, some schemes perform considerably better than others. We demonstrate that, in selective security setting, the algorithm based on the post-quantum LWE achieves better performance than DDH in terms of time overhead. However, in the adaptive security setting, LWE is hardly usable in memory-constrained application scenarios.

\section{Background}\label{sec:background}
We present two notions of security we based our schemes on. A detailed description is provided in  \cite{fromseltoadpt} and  \cite{fully_secure}.

\emph{Adaptive Security}. We say that a scheme is \textit{adaptively secure} if it is resistant against 
adaptive attacks, i.e. where the attacker is allowed to choose the pair of messages in the challenge phase, based on the previously collected
information.

\emph{Selective Security}. We say that a scheme is \textit{selectively secure} if it is resistant against selective
attacks, i.e. where the attacker has to declare the pair of challenge messages at
the outset of the game (that is, before seeing the master public key). In this setting, there is no previous information the attacker can use. 

Of course, the adaptive security notion is stronger than the selective one.
Still, adaptively secure schemes are less efficient than selectively secure ones.
The choice between these two levels of security comes off as a classic evaluation of the tradeoff between security and performance.

We now present the main cryptographic model we use. \textit{\acl{fe}} \cite{Mascia_Sala_Villa_23} is an extended form of the public-key setting, where a functional key $sk_f$ can be derived from a master secret key $msk$ for a certain function $f$. By applying $sk_f$ to the encryption of a plaintext $x$, $f(x)$ will be revealed without decrypting the ciphertext.

A single-input \textit{\ac{ipfe}} is a type of \ac{fe} that supports the evaluation of inner product on the encrypted data. Given the encryption of a plaintext vector $\mathbf{x} = \extVec{x}{m}$, and a functional key $sk_\mathbf{y}$ linked with a vector $\mathbf{y}=\extVec{y}{m}$, the decryption function, run with $sk_\mathbf{y}$, outputs $\scalProd{x}{y} = \scalProdRes{x}{y}{m}$.

The single-input \ac{ipfe} can be lifted in a \textit{multi-input} (MI) setting where different vectorial plaintexts $\mathbf{x}_1,\dots,\mathbf{x}_n\in\group{q}^m$ are encrypted with different keys and in the decryption phase, using a functional key $sk_\mathbf{y}$ linked to a vector $\mathbf{y} = (\mathbf{y}_1,\dots,\mathbf{y}_n)$, the inner product (\Cref{eq:inner_prod}) will be revealed.
\begin{equation} 
    \label{eq:inner_prod}
    \sum_{i = 1}^n{\langle {\bf x}_i,{\bf y}_i\rangle}
\end{equation}

The IPFE schemes are composed by four algorithms. A \textsc{Setup} phase in which the public parameters, the master secret key and the clients' private keys are generated. A \textsc{KeyGen} phase in which the functional key, linked with a vector $\mathbf{y}$, is derived from the master secret key. An \textsc{Encryption} phase where the plaintexts are encrypted and a \textsc{Decryption} phase where the inner product is computed using the functional key.\\
We have selected two classes of \ac{mife} schemes for inner product: the first one bases its security on the \textit{\acf{ddh}} problem, while the second is constructed over the \textit{\acf{lwe}} problem. While the first one is already used in some \ac{fl} implementations \cite{xu_hybridalpha_2019}, we have not found any benchmark in the second setting, at the best of our knowledge, even if it is a post-quantum solution.

\begin{algorithm}[t!] 
 \caption{Description of the DDH-based single-input \ac{ipfe} scheme
 in \textcolor{cb_cyan}{selective} (left) and \textcolor{cb_caramel}{adaptive} (right) secure setting.
}
\label{alg : DDH}
    \begin{algorithmic}
        \Procedure{Setup}{$1^\lambda, m$}
        \State $(\mathcal{G}, q, \text{g}) \gets \textsc{GroupGen}(1^\lambda)$\\
        \vspace{0.1cm}
        \hspace{0.23cm}
        \begin{tabularx}{\linewidth}{>{\raggedright\arraybackslash}p{0.375\linewidth} | >{\raggedright\arraybackslash}X}
            \textcolor{cb_cyan}{$\mathbf{s} \gets_R \group{q}^m$}  &
            \textcolor{cb_caramel}{$a \leftarrow_R \group{q},\ \mathbf{a}=  (1, a )^\top$} \\

            \textcolor{cb_cyan}{$\mathbf{h} \gets (\text{g}^{s_1}, \dots, {\text{g}}^{s_m})$}  &
            \textcolor{cb_caramel}{$\textbf{W} \longleftarrow_R \group{q}^{m \times 2}$} \\

            $\textcolor{cb_cyan}{\mathit{msk} \gets \mathbf{s}, \ \mathit{mpk} \gets \mathbf{h}}$  &
            $\textcolor{cb_caramel}{\mathit{msk} \gets \textbf{W}, \ \mathit{mpk} \gets (\text{g}^\mathbf{a}, \text{g}^{\textbf{W}\mathbf{a}})}$ \\
        \end{tabularx}
        \State \Return $(\mathit{msk}, \mathit{mpk})$
        \EndProcedure
        \Procedure{Enc}{$mpk, \mathbf{x} \in \group{q}^m$}
         \State $r \gets_R \group{q}$\\
        \vspace{0.1cm}
        \hspace{0.23cm}
        \begin{tabularx}{\linewidth}{>{\raggedright\arraybackslash}p{0.375\linewidth} | >{\raggedright\arraybackslash}X}
        \textcolor{cb_cyan}{$ct_0 \gets \text{g}^r$} &  \textcolor{cb_caramel}{$\mathbf{ct}' \gets \text{g}^{\mathbf{a}r}$}\\
         \textcolor{cb_cyan}{$\forall i\in [m] \  ct_i \gets h_i^r \ \text{g}^{x_i}$} &  \textcolor{cb_caramel}{$\mathbf{ct}'' \gets \text{g}^{\mathbf{x} +\mathbf{W}\mathbf{a}{r}}$}\\
         \textcolor{cb_cyan}{$\mathbf{ct} \gets (ct_0, (ct_i)_{i \in [m]})$} & \textcolor{cb_caramel}{$\mathbf{ct} \gets (\mathbf{ct}',\mathbf{ct}'')$}\\
         \end{tabularx}
        \State \Return $\mathbf{ct}$
    \EndProcedure
    \Procedure{KeyGen}{$msk, \mathbf{y} \in \group{q}^m$}\\
        \vspace{0.1cm}
        \hspace{0.23cm}
        \begin{tabularx}{\linewidth}{>{\raggedright\arraybackslash}p{0.375\linewidth} | >{\raggedright\arraybackslash}X}
         \textcolor{cb_cyan}{$sk_{\mathbf{y}} \gets (\langle \mathbf{y}, \mathbf{s}\rangle, \ \mathbf{y})$} & \textcolor{cb_caramel}{$sk_{\mathbf{y}} \gets ( \ \mathbf{W}^\top \mathbf{y}, \mathbf{y})$}
        \end{tabularx}
        \State \Return $sk_{\mathbf{y}}$
    \EndProcedure
    \Procedure{Dec}{$\mathbf{ct},\ sk_{\mathbf{y}} = (d, \mathbf{y})$}\\
        \vspace{0.1cm}
        \hspace{0.23cm}
        \begin{tabularx}{\linewidth}{>{\raggedright\arraybackslash}p{0.375\linewidth} | >{\raggedright\arraybackslash}X}
         \textcolor{cb_cyan}{$C \gets \dfrac{\prod_{i=1}^{m} ct_i^{y_i}}{ct_0^d}$}  & \textcolor{cb_caramel}{$C \gets \frac{\prod_{i =1}^m ({ct''}_i)^{y_i}} {\prod_{i=1}^2 ({ct'}_i)^{d_i}}$}\\
        \end{tabularx}
        \State $res \gets \textsc{Dlog}_{\text{g}}(C)$
        \State \Return $res$
    \EndProcedure
    \end{algorithmic}
\end{algorithm}

In both cases, in order to design a multi-input scheme we start from a single-input scheme and lift it to a multi-input setting using the compiler presented in \cite{noPairing}.
To cover a wider range of use cases, we use both a \Cb{selective-secure} and an \Cr{adaptive-secure} \cite{adapt_sel_SI} single-input scheme for each problem, thus obtaining different schemes with varying properties.
In the following, we are going to use the colors above to stress the differences in the \Cb{selective} and \Cr{adaptive} case. Moreover, in algorithms description, we make use of the bracket notation $[n]$ to denote the set $\{1, \dots, n\}$.
\subsubsection{DDH-based FE}
both the selective-secure and the adaptive-secure scheme are based on the plain DDH problem. The first one is derived from the ElGamal PKE and translated into an IPFE scheme \cite{katz_simple_2015}, while the second is discussed in detail in \cite{noPairing}.
 Let \textsc{GroupGen} be a probabilistic polynomial time algorithm that takes as input a security parameter $1^\lambda$ and outputs $(\mathcal{G},\text{q}, \text{g})$, where $\mathcal{G}$ is a group of prime order $\text{q}$ generated by \text{g}. \Cref{alg : DDH} briefly describes the schemes.

\begin{algorithm}[t!] 
\caption{Description of the LWE-based single-input \ac{ipfe} scheme in \textcolor{cb_cyan}{selective} (left) and \textcolor{cb_caramel}{adaptive} (right) secure setting.
}
\label{alg: LWE}
    \begin{algorithmic}
        \Procedure{Setup}{$1^\lambda, m$}
        \State $\mathbf{A} \longleftarrow_R \group{q}^{M\times N}$\\ 
        \vspace{0.1cm}
        \hspace{0.23cm}
        \begin{tabularx}{\linewidth}{>{\raggedright\arraybackslash}p{0.4\linewidth} | >{\raggedright\arraybackslash}X}
        \textcolor{cb_cyan}{ $\mathbf{S} \longleftarrow_R \group{q}^{N\times m}$} & \textcolor{cb_caramel}{$ \mathbf{S} \leftarrow_R \mathcal{D}$} \\
        \textcolor{cb_cyan}{$\mathbf{E} \longleftarrow_R \mathcal{X}_{\bar{\sigma}}^{M \times m}$}\\
        \textcolor{cb_cyan}{$\mathbf{U} \gets \mathbf{A} \mathbf{S}+ \mathbf{E} $} & \textcolor{cb_caramel}{$ \mathbf{U} \gets  \mathbf{S} \mathbf{A}$} 
        \end{tabularx}
        \State $mpk \gets (\mathbf{A},\mathbf{U}), \ msk \gets \mathbf{S}$
        \State \Return $(\mathit{msk}, \mathit{mpk})$
    \EndProcedure
    \Procedure{Enc}{$mpk, \mathbf{x} \in \group{}^m$}\\
        \vspace{0.1cm}
        \hspace{0.23cm}
        \begin{tabularx}{\linewidth}{>{\raggedright\arraybackslash}p{0.4\linewidth} | >{\raggedright\arraybackslash}X}
        \textcolor{cb_cyan}{$\mathbf{r} \longleftarrow_R \{ 0,1 \}^M$} & \textcolor{cb_caramel}{$ \mathbf{s} \leftarrow_R \group{q}^{N}$}\\
         & \textcolor{cb_caramel}{$ \mathbf{e}_0 \leftarrow_R \mathcal{X}_{\alpha q}^{M} \ \mathbf{e}_1 \leftarrow_R \mathcal{X}_{\alpha q}^{m}$}\\
           \textcolor{cb_cyan}{$\mathbf{ct}{'} \gets \mathbf{A}^\top \mathbf{r}$} & \textcolor{cb_caramel}{$\mathbf{ct}{'} \gets \mathbf{A}\mathbf{s} + \mathbf{e}_0$} \\
           \textcolor{cb_cyan}{$\mathbf{ct}{''} \gets \mathbf{U}^\top \mathbf{r} + t(\mathbf{x})$}  & \textcolor{cb_caramel}{$\mathbf{ct}{''} \gets {\mathbf{U} \mathbf{s} + \mathbf{e}_1 + \mathbf{x} \cdot \lfloor \frac{q}{K} \rfloor}$}
        \end{tabularx}
        \State \Return  $\mathbf{ct} \gets (\mathbf{ct}{'}, \mathbf{ct}{''} )$
    \EndProcedure
    \Procedure{KeyGen}{$msk, \mathbf{y} \in \group{}^m$}\\
        \vspace{0.1cm}
        \hspace{0.23cm}
        \begin{tabularx}{\linewidth}{>{\raggedright\arraybackslash}p{0.4\linewidth} | >{\raggedright\arraybackslash}X}
        \textcolor{cb_cyan}{$sk_\mathbf{y} \gets ({msk \cdot \mathbf{y}}, \mathbf{y})$} &  \textcolor{cb_caramel}{$sk_\mathbf{y} \gets ({msk^\top \mathbf{y}}, \mathbf{y})$}
        \end{tabularx}
        \State \Return $sk_{\mathbf{y}}$
    \EndProcedure
    \Procedure{Dec}{$\mathbf{ct},\ sk_{\mathbf{y}} = (\mathbf{d}, \mathbf{y})$}
        \State $C \gets \mathbf{y} \cdot \mathbf{ct}'' - \mathbf{d} \cdot \mathbf{ct}' \mod{q}$
        \State $res \longleftarrow \text{the plaintext }  x  \text{ that minimizes:}$ \\
        \vspace{0.1cm}
        \hspace{0.23cm}
        \begin{tabularx}{\linewidth}{>{\raggedright\arraybackslash}p{0.4\linewidth} | >{\raggedright\arraybackslash}X}
        $ \textcolor{cb_cyan}{\vert C-t(x) \vert } $ & $ \textcolor{cb_caramel}{\vert \left\lfloor \frac{q}{K} x \right\rfloor - C \vert} $ \\
        \end{tabularx}
        \State \Return $res$
    \EndProcedure
    \end{algorithmic}
\end{algorithm}

\subsubsection{LWE-based FE}
the LWE-based \ac{ipfe} schemes that we selected are \textit{Bounded-Norm Inner Product} schemes. This means that each plaintext $\mathbf{x}$ and vector $\mathbf{y}$ has to be respectively such that $\|\mathbf{x}\|_\infty < X$ and $\|\mathbf{y}\|_\infty < Y$ for some fixed $X,Y$.

The single-input selective-secure scheme is derived from Regev PKE and turned into a IPFE scheme as explained in \cite{katz_simple_2015}, while the adaptive-secure single-input scheme is presented in \cite{noPairing}. The latter bases its security on a variant of the LWE problem called the \textit{multi-hint extended-LWE} (mheLWE) problem, which is not easier than the plain LWE problem, as there is a reduction from LWE to mheLWE \cite{fully_secure}.

A brief description of both adaptive and selective schemes is presented in \Cref{alg: LWE}. In the pseudocode, $m$ denotes the length of an input vector, while $M, N$ are security parameters for LWE. With reference to \Cref{alg: LWE}, $M$ is the dimension of the ciphertext expansion through additional random generated LWE instances (i.e. $ct'$), while $N$ is the dimension of the instance itself, that is, the dimension of the LWE secret $s$ that we use as a mask. In general, $M = \Theta(N \log q)$ (see \cite{fully_secure}).
$\mathcal{X}_{\bar{\sigma}}$ denotes an integer Gaussian distribution over $\group{q}$ with standard deviation $\bar{\sigma}$ and $\mathcal{D}$ is a distribution over $\group{}^{m \times M}$ as defined in \cite{fully_secure}. Given two primes $p,q$ with $q > p$, for every $v \in \group{p}$ let the center function be defined as $t(v) = \lfloor v \cdot \frac{q}{p} \rfloor \in \group{q}$ and $\alpha $ a real number in $(0,1)$.

\subsubsection{From Single-Input to Multi-Input}
in order to lift the presented schemes to a multi-input setting, we used the compiler proposed by M. Abdalla et al. \cite{noPairing} and summarized in \Cref{alg:compiler}. This compiler works when the single-input FE scheme satisfies two properties 
called \textit{Two-step decryption} and \textit{Linear encryption}.

\begin{algorithm}[t!] 
\caption{$\mathcal{(MIFE)}$ Compiler of \cite{noPairing} that lifts a single-input IPFE $\mathcal{FE}$ to a multi-input setting.}
\label{alg:compiler}
\begin{algorithmic}
    \Procedure{Setup}{$1^\lambda, m, n$}
        \ForAll{$i \in [n]$}
            \State $\mathbf{u}_i \longleftarrow_R \group{q}^m$
            \State $(msk_i', mpk_i') \longleftarrow \mathcal{FE}\text{.Setup}(1^\lambda, m)$
            \State $csk_i \gets (mpk_i', \mathbf{u}_i)$
         \EndFor
        \State $msk \gets ((msk_i')_i, (\mathbf{u}_i)_i)$
        \State \Return $(msk, (csk_i)_i)$
    \EndProcedure
    \Procedure{Enc}{$csk_i, \ \mathbf{x}_i \in \group{q}^m$}
        \State $ct_i \longleftarrow \mathcal{FE}.\text{Enc}(mpk_i', \mathbf{x}_i + \mathbf{u}_i)$
        \State \Return $(ct_i)_{i \in [n]}$
    \EndProcedure
    \Procedure{KeyGen}{$msk, \ \mathbf{y}: \mathbf{y} = (\mathbf{y}_i)_{i \in [n]},\ \mathbf{y}_i \in \group{q}^m$}
        \ForAll{$i \in [n]$}
            \State $sk_{i,\mathbf{y}} \longleftarrow \mathcal{FE}.\text{KeyGen}(msk_i', \mathbf{y}_i)$
        \EndFor
        \State $z \gets \sum_{i \in [n]} \langle \mathbf{u}_i, \mathbf{y}_i\rangle \in \group{q}$
        \State $sk_\mathbf{y} \gets ((sk_{i,\mathbf{y}})_i, z)$
        \State \Return $sk_\mathbf{y}$
    \EndProcedure
    \Procedure{Dec}{$sk_\mathbf{y},\ ct_1,\dots,ct_n$}
        \ForAll{$i \in [n]$}
            \State $D_{i,1} \longleftarrow \text{Dec}_1(ct_i, sk_{i,\mathbf{y}})$
        \EndFor
        \State $res \longleftarrow \text{Dec}_2\Bigl( \sum_{i\in [n]} D_{i,1}, z\Bigr)$
        \State \Return $res$
    \EndProcedure
\end{algorithmic}
\end{algorithm}

When \acl{fl} is combined with MIFE, an additional entity is often introduced: the \textit{\ac{tpa}}. The \ac{tpa} is responsible for setup, key generation, and key distribution, while the \ac{fl} server performs the decription.\\

\section{Related work} \label{sec:related}
One of the main concerns in \ac{fl} is the risk of data leakage due to various types of attacks. This privacy concern can be classified into two main categories\cite{nist_federated_privacy}: privacy of the local model and privacy of the output model. Privacy of the local model means that no one, including the server, should have access to updates of the individual clients model. Privacy of the output model means that no one can extract information about the training data from the model computed by the server at each round.
A scheme that satisfies both privacy requirements is called \textit{Privacy-Preserving Federated Learning (PPFL)}~\cite{chang2023privacy}.

To obtain privacy of the local model some cryptographic primitives can be used. The main directions are represented by \textit{\ac{he}}, \textit{\ac{mpc}} with \textit{Secret Sharing} techniques and {\ac{fe}}.
In this work we focus on MIFE.

Regarding the privacy of the output model, a well known approach is  \textit{Differential Privacy (DP)} \cite{nist_federated_privacy}: a non-cryptographic mechanism that consists in adding some noise to the information that are sent to the server. A widely popular algorithm, that adds
 noise during the training process, is the \textit{Differential Private Stochastic Gradient Descent (DP-SGD)}.
 
\textit{HybridAlpha} \cite{xu_hybridalpha_2019} represents the first application of \textit{\ac{mife}} to \ac{fl}. The involved entities are the Server, the Clients and the TPA that handles key generation and distribution. This protocol uses DP and MIFE to solve some privacy concerns, but leaving some security issues (see \Cref{sec:threat}) regarding the privacy of the local model. Building on the HybridAlpha framework, other works have explored the developing of PPFL using DP mechanisms and MIFE \cite{ppfl_iot,xu2021fedv,qian2022cryptofe}.

Other solutions have moved towards a decentralized setting in order to not rely on a TPA \cite{chang2023privacy}. The main problem of these solutions is the need of cross-client interactions, which result in additional communication overhead and poor scalability.
\section{Threat model}\label{sec:threat}
We consider a \ac{fl} scenario with an honest-but-curious server that follows the correct procedures for parameter aggregation and client selection, but that may attempt to infer clients' confidential information by exploiting the knowledge shared by clients during the federated training process. This information includes the model's parameters and the global model's accuracy on local validation sets (in the case of \ac{flad}).
Using this information, along with knowledge of the global model's architecture, the server could attempt to reconstruct private information from the clients' training data \cite{10.5555/3495724.3497145}.

We assume that a subset of clients may be dishonest and collude to obtain private information from other clients (e.g. they could share their model to infer information on the other clients' parameters). Throughout the training process, we suppose that at least two clients among the participants to the protocol are honest. 
Moreover, we assume that the clients do not have the ability to manipulate the training data to compromise the global model's operations. This type of attack, known as poisoning attack, is a well-known problem \cite{wan2024data,tian2022comprehensive}, but it is outside the scope of this work. 
Finally, the \ac{tpa}, which is responsible for distributing the keys for \ac{mife} and the signatures needed to authenticate communications, is assumed to be honest by both the clients and the server.

Our work is based on HybridAlpha framework. HybridAlpha suffers from some security issues \cite{chang2023privacy}, in particular, there exist two attacks that lead the server to obtain more
information about the clients' model than what it should.

The first attack, referred to as \textit{ciphertexts mix-and-match},
decrypts sums of ciphertexts from different rounds to obtain other information \cite{chang2023privacy}.


The second attack, called \textit{decryption-
keys mix-and-match}, allows the server to retrieve the model
parameters sent by a client using different decryption keys
from different rounds.

To mitigate these leakages, the \textit{Multi-RoundSecAgg} framework was introduced in~\cite{so2023securing}. This solution presents a trade-off between privacy and convergence time of the global model. 
Additional solutions have been proposed in a decentralized setting, as in \cite{chang2023privacy}. Our scheme addresses these data leakages without using these approaches.

\section{Challenges}\label{sec:features}

In this section, we highlight the features of \ac{fl} that are critical from a security standpoint, with respect to our threat model. Specifically, we focus on a recent state-of-the-art \ac{fl} approach for training \ac{dl}-based \acp{nids} called \ac{flad} \cite{flad}. Compared to the standard \ac{fl} process, \ac{flad} introduces a mechanism in which participants share the accuracy score of the global model, as measured on their local validation sets, with the \ac{fl} server. This approach enables \ac{flad} to outperform \acs{fvg}, the model at the core of FL \cite{mcmahan2017communication}, in both accuracy and training efficiency on unbalanced and heterogeneous datasets of network intrusions, thus making \ac{flad} the state-of-the-art in its field of application.


\textit{Aggregation}. In FLAD, aggregation is performed by the server computing the arithmetic mean of the clients' parameters, regardless of whether they trained or not in the current round. This step is performed in clear, revealing the partial client's models to the server and allowing it to perform reconstruction attacks, partially leaking the client's private training dataset.

\textit{Client selection and steps/epochs assignment}. In FLAD, at the beginning of each round, the server selects the set of clients that have to train their local model and adaptively assigns them a number of steps and epochs. The client selection and the steps and epochs assignment procedures are functions of the mean accuracy $a^{\mu}$ and the local model accuracy $a^i$ for each client $i$. Thus, these quantities need to be known by the server. This knowledge leaks which subset of clients is involved in the training process at the current round, allowing the server to deduce information about a target client's model. For example, in the case where only a client is training in some round, the server can deduce its model from the aggregated one.

\textit{Termination criterion}. FLAD makes use of a patience-based termination criterion, which also depends on $a^{\mu}$, thus, indirectly, on the clients' scores, incurring in the same leakage of the previous point.
\vspace{.2cm}

We address all these critical features by taking advantage of the MIFE properties. In particular, as we describe in depth in \Cref{sec:methodology}, any sensitive information about a client, such as the model and the scores, is encrypted with MIFE before being communicated to the server. Hence, the server only gets to know the aggregated values. 

Moreover, we moved the steps and epochs assignment procedure on the client side, in order not to leak to the server the subset of participants involved in a training round.
\vspace{.2cm}
\begin{remark}
    The original FLAD procedure for assigning steps and epochs is also a function of the minimum and maximum of the set of scores. After MIFE aggregation, the clients only have access to the mean score $a^{\mu}$, thus, we slightly modified this procedure, as presented in \Cref{sec:methodology}. As the tests remarked, this modification have no particular impact on the convergence of the training in terms of rounds.
\end{remark}
\section{Scheme description}\label{sec:methodology}
\begin{figure}
    \centering
\includegraphics[width=0.495\textwidth]{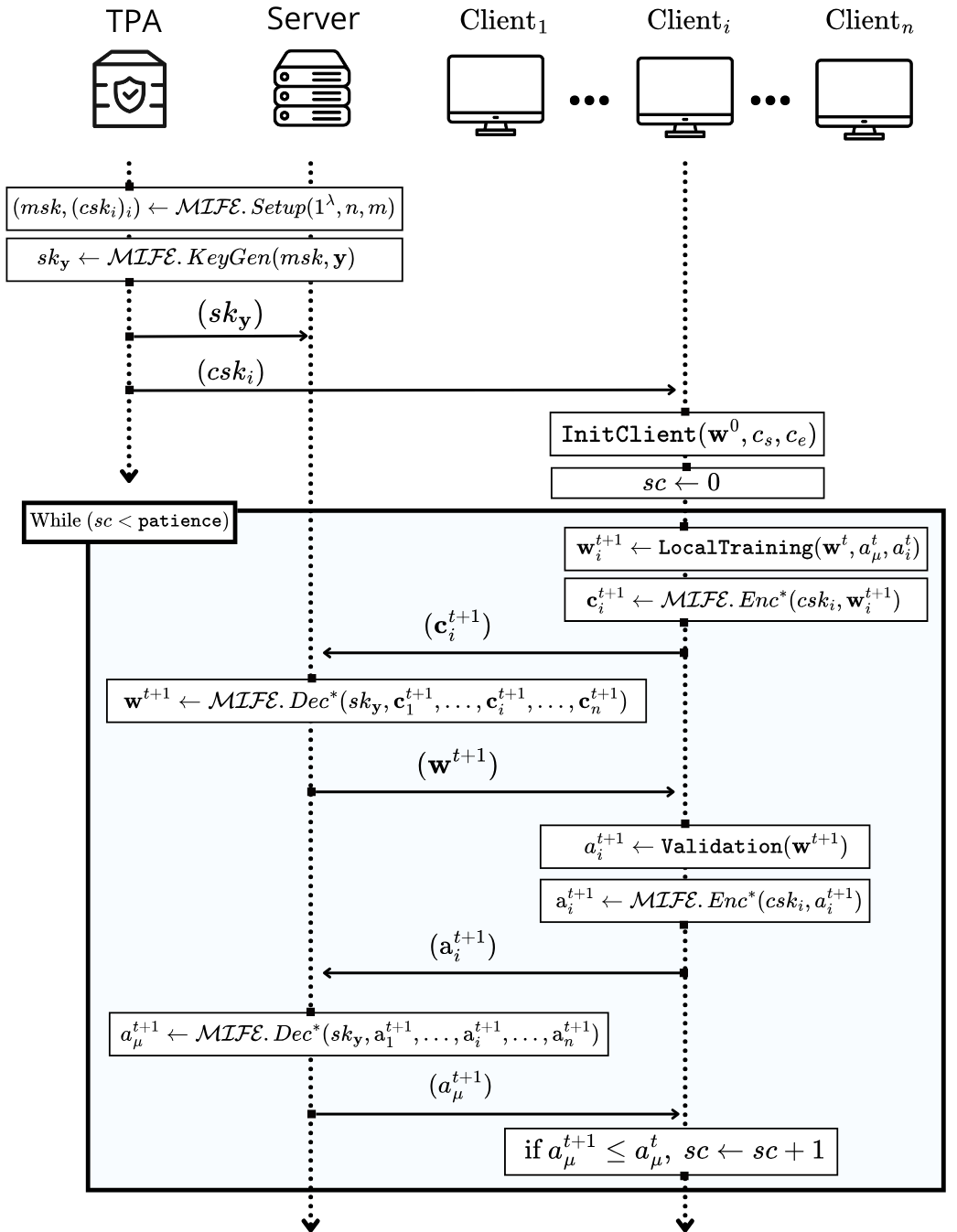}
    \caption{Protocol description.}
    \label{fig:full_protocol}
\vspace{-3mm}
\end{figure}

In this section, we present a description of our proposed scheme,
 depicted in \Cref{fig:full_protocol}. It extends the original FLAD protocol \cite{flad} by incorporating a security layer based on MIFE.
Here we abstract from a specific MIFE instantiation. 

First, the \ac{tpa} performs the setup and generates the keys. Then, it distributes to each of the $n$ clients their secret key $csk_i$ and a common seed $s_0$. This seed, along with a previously agreed collision resistant pseudo-random function $f$, is used to generate some labels $\gamma_t:=f(s_0,t)$.
The label $\gamma_t$ is added to each plaintext in the $t$-th round before the real encryption, as shown in \Cref{eq:labels}. This prevents the server from performing a \textit{ciphertext mix-and-match} attack, since the output of a decryption is now masked by the labels.

The \ac{tpa} provides the server with the decryption key $sk_\mathbf{y}$ associated with the vector $\mathbf{y} = (1,\dots, 1)\in\mathbb{Z}^n$. This is the only key the server needs, since in each round it has to compute the sum of the clients' parameters and scores.
\begin{remark}
\label{rem:shape_y}
   a constant decryption key throughout the rounds prevents a \textit{decryption-key mix-and-match attack}, since the server has only access to one key. For the decryption key to be constant, we require all clients to participate in each round, regardless of whether they have trained in the current round or not. If they did not train, they must encrypt the aggregated model of the previous step, with a new randomness and a label consistent with the current round.
\end{remark}
\vspace{.1cm}
In our implementation, the parameter $m$, which represents the number of entries in each plaintext vector that we want to sum in each MIFE decryption, is set to $1$. This is because the server has to compute sums across the corresponding entries of the input vectors of different clients. In fact, with $m>1$, the server would compute a single sum among the first $m$ entries of all client's own input, consistent with Equation \ref{eq:inner_prod}, resulting in an output that is meaningless for our purposes.

At this point, the clients run the algorithm \textsc{InitClients($\omega_0, c_s,c_e$)} from the original FLAD scheme (\cite{flad}). This means that every client initializes its own model, and for the first round everyone trains with the maximum amount of steps and epochs.

In each round, every client either trains its model or not, depending on the personal score obtained in the previous round. Regardless of whether training was performed, the client encrypts its parameters and sends them to the server. This is a key point of our scheme, as it allows the decryption key to be constant, as anticipated in \Cref{rem:shape_y}.

In the parameters encryption phase, at the $t$-th round, first of all the $i$-th client encodes the real valued parameters ${\bf w}^t_i = (w_{1,i}^t,\dots,w_{l,i}^t )$ in integers, by setting a precision $\Delta$ shared by each client and computing
\begin{equation} \footnotesize
\label{eq:encoding}
\begin{pmatrix}
x_{1,i}^t \\
\vdots \\
x_{l,i}^t 
\end{pmatrix}
:=
\begin{pmatrix}
[10^\Delta\cdot w_{1,i}^t] \\
\vdots \\
[10^\Delta\cdot w_{l,i}^t] 
\end{pmatrix}.
\end{equation}
Note that, once this precision $\Delta$ is set, MIFE works in exact arithmetic; thus, it does not affect the convergence of the training or the parameters of the models. Theoretically, one could set $\Delta$ to machine precision, with a trade-off between accuracy and efficiency that will be discussed in \Cref{sec:results}.

Then, the client computes the label $\gamma_t$, adds it to each plaintext $(x_{1,i}^t,\dots,x_{l,i}^t)$ obtaining
\begin{equation} \footnotesize
\label{eq:labels}
\begin{pmatrix}
e_{1,i}^t \\
\vdots \\
e_{l,i}^t 
\end{pmatrix}
:=
\begin{pmatrix}
x_{1,i}^t \\
\vdots \\
x_{l,i}^t 
\end{pmatrix}
+ \begin{pmatrix}
\gamma_t \\
\vdots\\
\gamma_t \\
\end{pmatrix}.
\end{equation}
At this point $\mathcal{MIFE}.Enc$ is computed separately on each $e_{j,i}^t$,
\begin{equation} \footnotesize
\label{eq:encyrption}
{\bf c}^t_i =
\begin{pmatrix}
 c_{1,i}^t \\
\vdots \\
c_{l,i}^t
\end{pmatrix}
:=
\begin{pmatrix}
 \mathcal{MIFE}.Enc(csk_i,e_{1,i}^t)\\
\vdots \\
  \mathcal{MIFE}.Enc(csk_i,e_{l,i}^t)
\end{pmatrix}.
\end{equation}
This step is done without batching, meaning that each parameter is encrypted individually.

After that the server has received the ciphertexts from all clients, it computes $l$ $\mathcal{MIFE}.Dec$ on inputs $(c_{j,1}^t,\dots,c_{j,n}^t)$ for each $j$ $ \in [l]$
\begin{equation} \footnotesize
\label{eq:decryption}
\begin{pmatrix}
r_1^t \\
\vdots \\
r_l^t
\end{pmatrix}
:=
\begin{pmatrix}
 \mathcal{MIFE}.Dec(sk_\mathbf{y},c_{1,1}^t,\dots,c_{1,n}^t)\\
\vdots \\ \mathcal{MIFE}.Dec(sk_\mathbf{y},c_{l,1}^t,\dots,c_{l,n}^t)
\end{pmatrix}.
\end{equation}
In this way it obtains the sums of the parameters along with $n$ times the label that identifies the round.

The server sends this result to the clients, that subtract $n$ times the round label
\begin{equation} \footnotesize
\label{eq:delabeling}
\begin{pmatrix}
m_1^t \\
\vdots \\
m_l^t
\end{pmatrix}
:=
\begin{pmatrix}
{r_1^t-n\cdot\gamma_t }\\
\vdots \\
{r_l^t-n\cdot\gamma_t}
\end{pmatrix}.
\end{equation}
Finally, the client maps this integer back to floating points and perform a division by $n$ to compute the means of $(w_{j,1}^t,\dots,w_{j,n}^t)$ for all $j$, that is:
\begin{equation} \footnotesize
\label{eq:decoding}
{\bf w}^{t}_i=
\begin{pmatrix}
w_1^{t} \\
\vdots \\
w_l^{t}
\end{pmatrix}
:=
\begin{pmatrix}
\frac{m_1^t}{10^\Delta\cdot n }\\
\vdots \\
\frac{m_l^t}{10^\Delta\cdot n }
\end{pmatrix},
\end{equation}
and uses them to obtain accuracy scores of this new aggregated model on their own validation set.


Then, the clients encrypt their score using the same procedure followed for parameters' encryption and send it to the server. The server computes $\mathcal{MIFE}.Dec$ and returns the result to the clients, which can now compute the mean accuracy score value $a^{\mu}$. At this point, each client runs \Cref{alg: trainload} on inputs its score $a^i$ and $a^{\mu}$, to obtain the number of epochs $c_e$ and steps $c_s$ of training for the next round. In \Cref{alg: trainload}, the quantities $e_{min}$, $e_{max}$, $s_{min}$ and $s_{max}$ are constants, representing the minimal and maximal number of epochs and steps we want the clients to train for.
\begin{algorithm}
\caption{Assigment of steps and epochs.}
\label{alg: trainload}
\begin{algorithmic}
\Procedure{TrainingLoad}{$a^i$, $a^\mu$}
    \If{$a^i \leq a^{\mu}$}
        \State $\sigma =\lvert \frac{a^{\mu}-a^i}{a^{\mu}} \rvert$ \Comment{Scaling factor}
        \State $c_e = e_{\min} + (e_{\max} - e_{\min}) \cdot \sigma$
        \State $c_s = s_{\min} + (s_{\max} - s_{\min}) \cdot \sigma$
    \Else
        \State $c_e = 0$, $c_s =0$
    \EndIf
\EndProcedure
\end{algorithmic}
\end{algorithm}

The assignment of steps and epochs is summarized in \Cref{fig:full_protocol} in the invocation to the method \texttt{LocalTraining}().
We assume that during the training process each client uses a differential privacy mechanism as in \cite{xu_hybridalpha_2019}. We will not dive further into it as it is out of scope for this work. 

\Cref{fig:full_protocol} proposes a simplified brief description of the overall process. $\mathcal{MIFE}.Enc^*$ and $\mathcal{MIFE}.Dec^*$ denote respectively encryption and decryption algorithm of the $\mathcal{MIFE}$ scheme over the vector of parameters without batching. For the sake of clarity, labeling and encoding were not included in the figure, see \Cref{eq:encoding,eq:labels,eq:encyrption}for details. Note also that the decryption is actually composed of a server-side step, that is \Cref{eq:decryption} and a client-side step, namely Equations (\ref{eq:delabeling}) and (\ref{eq:decoding}). For the sake of simplicity, we ignored these details in \Cref{fig:full_protocol}, as they are discussed in detail in this section.

\subsection{Termination Criterion}
The FL process terminates following the same procedure of \cite{flad}. The difference is that the decision process is moved to the client side, as the server does not know the mean and maximum score. When the process concludes, the clients report it to the server. Some dishonest clients may not agree with the honest ones on the termination. To address such problem and also detect which clients are dishonest, we also require the server to keep a log of the last $r$ aggregated mean scores, where $r$ = \textit{patience}. In this way, whenever the clients disagree on termination, the \ac{tpa} will be able to identify the dishonest clients by consulting the log.


\subsection{Joining and dropouts of clients}
Our scheme also supports clients to  join and dropout.
Let $n$ be the number of clients currently participating to the process and $t$ the current round. These are public parameters known by all entities.

As observed in \Cref{sec:threat}, in each round, at least $2$ clients have to be honest. Consequently, if we allow up to $s$ drop out, we need to start with at least $s+2$ honest clients.

When a client joins, it receives the private key and the seed by the TPA. The latter updates the decryption key $sk_\mathbf{y}$ and sends it to the server.  However, it should be noted that associating a new $u_i$ to the new client exposes our scheme to a \textit{mix-and-match} attack (\Cref{sec:threat}) on the updated $sk_\mathbf{y}$. In fact, this new $u_i$ can be directly obtained by subtracting $z$ in the old $sk_\mathbf{y}$ from the new $z$. The simplest way to avoid this is to change the $u_i$ of a random client when a new client joins. 

If a client $i_0$ drops out the process, for the scheme to keep on working with the remaining subset of clients, it suffices to update $sk_\mathbf{y}$, by removing the term depending on $u_{i_0}$. Even in this case, it is necessary to change a random $u_i$ among the remaining clients to avoid a \textit{mix-and-match} attack.

This resulting scheme does not have the privacy leakages of HybridAlpha, that are highlighted in \cite{chang2023privacy}. Indeed, an honest-but-curious server can not obtain more information about the clients than an honest server would: the \textit{ciphertexts mix-and-match} attack is not applicable due to the usage of the labels. acting as a round-dependent one-time pad.
In addition, the \textit{decryption-keys mix-and-match} attack is infeasible, since the decryption key is the same in all rounds. The situation in which new clients join or drop out the training process is more critical. In fact, when the number of users varies, the private keys of individual clients remain unchanged while the decryption key changes. However, the use of labels and the change of one client's $u_i$ protects our scheme from this attack.

\section{Experimental setup}\label{sec:setup}
In our experiments, we used \ac{flad} Release 1.0~\cite{flad-github} in a Python 3.9 environment with Tensorflow version 2.7.1. 
All experiments were performed on a server with an AMD EPYC 9454P 48-Core Processor and 128 GB of RAM.

\begin{figure}[t] \label{fig:digits}
    \centering
    \begin{tikzpicture}[scale = 0.9]
        \begin{axis}[
            height=4 cm,
            width=1\linewidth,
            xlabel={$\Delta$},
            ylabel={Final Accuracy},
            legend style={at={(0.6,1.1)}, anchor=south west, draw=none, legend columns=-1},
            axis y line*=left, 
            ymin=0.97, ymax=0.99, 
            ytick={0.970, 0.975, 0.980, 0.985, 0.990}, 
            yticklabel style={
            /pgf/number format/fixed,
            /pgf/number format/precision=3,
            /pgf/number format/fixed zerofill
            },
            xtick={1,2,3,4,5,6,7,8,9,10},
            ymajorgrids=true
        ]
        \addplot[color=Tomato Red, mark=*] table [x=Decimals, y=Last_Accuracy, col sep=comma] {sections/sources/averaged_results.csv};
        \addlegendentry{Final Accuracy}
        \end{axis}

        \begin{axis}[
            height=4 cm,
            width=1\linewidth,
            axis y line*=right, 
            ymin=34, ymax=52, 
            ylabel={Number of Rounds},
            legend style={at={(0,1.1)}, anchor=south west, draw=none, legend columns=-1}
            ]
        \addplot[color=Prussian Blue, mark=square*, dashed] table [x=Decimals, y=Number_of_Rounds, col sep=comma] {sections/sources/averaged_results.csv};
        \addlegendentry{Number of Rounds}
        \end{axis}
    \end{tikzpicture}
    \caption{Number of rounds and mean final accuracy over $10$ federated trainings with different random initial parameters for each decimal digit truncation $\Delta$.}
    \label{approx}
\vspace{-3mm}
\end{figure}

The experimental settings are consistent with those documented in the \ac{flad}'s paper \cite{flad} in terms of dataset (CIC-DDoS2019 \cite{cicddos2019}), data preprocessing (an array-like representation of traffic flows, where the rows of the grids represent packets in chronological order and columns are packet-level features), unbalanced and heterogeneous data across the clients (one and only one attack assigned to each client) and \ac{ml} model (a \ac{mlp} with two hidden layers of $32$ neurons each, and an output layer with a single neuron for binary classification of the network traffic as either benign or \acs{ddos}). The total number of parameters of this model is $4641$.

In order to achieve $128$ bits of security, we used Lattice Estimator \cite{lattice} to obtain LWE parameters that meet the bounds in \cite{fully_secure} Section 4.1, and NIST guidelines \cite{nist} for DDH parameters. The chosen parameters are shown in \Cref{tab:lwe_parameters}.

To minimize computational cost we approximate floating points to integers by multiplying them by a common factor $10^{\Delta}$, with $\Delta = 2$, and truncating the remaining digits. This choice is specific to our dataset. As shown in \Cref{approx}, $\Delta = 2$ is the most efficient approximation achieving both small bit length of plaintexts, leading to a faster MIFE, and good performance in federated training. 
We also exploit parallelization to speed up both the encryption and decryption of the models. We divided the list of plaintexts and ciphertexts into $15$ chunks of the same size, and assigned a different process to each chunk.

\begin{table}[t]
    \centering
    \begin{tabular}{l|cccc||l|c} 
        \toprule
        
         & $N$& $M$ & $log(q)$ & $\alpha$ &  & $log(q)$\\ 
        \midrule
        \multirow{2}{*}{LWE} &\Cb{80} & \Cb{$5327$} & \Cb{$63$} & \Cb{$1.09\times10^{-28}$}   & \multirow{2}{*}{DDH} & \Cb{3072}  \\
                             &\Cr{38} & \Cr{$9462$} &\Cr{$248$}  & \Cr{$1.71\times10^{-53}$}   &                       &\Cr{3072}  \\

        \bottomrule
    \end{tabular}
    \vspace{5pt}
    \caption{Parameters to achieve $\approx{128}$ bits of security.}
    \label{tab:lwe_parameters}
\vspace{-3mm}
\end{table}

\section{Experimental results}\label{sec:results}
We assess the memory and computational cost of the proposed scheme by running a full training. The primary goal of our tests is to evaluate the overhead introduced by the cryptographic layer on FLAD. Regardless of the cryptographic hardness assumption and the dataset, this overhead is linear in the number of training parameters. In fact, for both encryption and aggregation, each entry in the vector of model parameters undergoes the same operation individually. Consequently, conducting additional tests on different datasets would not provide significant new information.


In \Cref{tab:memory_cost}, we compare the proposed schemes by memory cost. This comparison does not take into account the impact of round labels $\gamma_t$. Note that, since the vector \textbf{y} consists entirely of ones, each entry requires only one bit for storage.

\begin{table}[b]
\centering \scriptsize
\renewcommand{\arraystretch}{1.2}
\begin{tabular}{P{0.5cm}|P{2.4cm}| P{2.4cm}| P{1.8cm}} \toprule

 & $\mathbf{\vert csk_i \vert}$ & $\mathbf{\vert sk_y \vert}$ & $\mathbf{\vert ct \vert}$ \\
 \midrule
\multirow{2}{*}{DDH}  & \Cb{$2 \log{(q)} $} & \Cb{$(n+1) \log{(q)}+n $} &   \Cb{$2 \log{(q)} $}\\
 & \Cr{$3 \log{(q)} $} & \Cr{$(2n+1) \log{(q)}+n$} &  \Cr{$3 \log{(q)} $ }\\
 \midrule
 \multirow{2}{*}{LWE}  & \Cb{$\log{(q)}(MN+M+1) $} & \Cb{$\log{(q)}(nN+1)+n$} &   \Cb{$\log{(q)}(N+1) $}\\
  & \Cr{$\log{(q)}(MN+N+1) $}& \Cr{$\log{(q)}(nM+1)+n$} &  \Cr{$\log{(q)}(M+1)$ }\\
\bottomrule
\end{tabular} 
    \caption{Memory cost comparison.}
    \label{tab:memory_cost}
\end{table}

Plugging the parameters shown in \Cref{tab:lwe_parameters} in the formulas in \Cref{tab:memory_cost}, a notable difference is the space required for each client's private key in DDH-based schemes compared to LWE-based ones. In particular, the formers require approximately $1$ KiB, whereas LWE-based schemes need $3$ to $10$ MiB. What stands out more about this comparison is the space required for each ciphertext in the adaptive LWE-based scheme, which is approximately $285$ KiB, whereas in the other three schemes it is around $1$ KiB. In our test, with $13$ clients and $4 \ 641$ parameters, we get that at each round the ciphertexts use around $16.5$ GiB of memory. This shows the poor scalability of this particular scheme compared to the others.

Figure \ref{fig:timebench} shows a comparison between the average time needed by the slowest client to train its model in a round of FLAD, the average time that the client uses to encrypt the model parameters and the aggregation time needed by the server to decrypt the parameters of the output model. 

Figure \ref{fig:timebench} shows that the encryption in DDH-based schemes is more efficient compared to LWE-based schemes. When considering the total time required for a full round, the selective LWE-based scheme outperforms the DDH-based scheme, while the adaptive LWE-based scheme is notably less efficient. However, LWE-based schemes offer quantum resistance, a feature not shared by DDH-based schemes.

\begin{figure}[t]
\centering
\begin{adjustbox}{max width=\linewidth}
\begin{tikzpicture}
\begin{axis}[
    ybar,
    bar width=25pt,
    width=0.9\textwidth,
    height=7cm,
    ylabel={Time (sec)},
    symbolic x coords={DDHselective, DDHadaptive, LWEselective, LWEadaptive},
    enlarge x limits=0.20,
    xtick=data,
    xtick style={draw=none},
    nodes near coords,
    nodes near coords align={vertical},
    legend style={at={(0.5,-0.15)},
        anchor=north,legend columns=-1},
    ylabel style={yshift=-1em},
    ymin=0, ymax=120,
    every node near coord/.append style={text=black} 
]
\addplot+[
    pattern=crosshatch,
    pattern color=cb_blue,
    draw=cb_blue]
    coordinates {(DDHselective,3.21) (DDHadaptive,3.19) (LWEselective,3.20) (LWEadaptive,3.30)};
\addplot+[
    pattern=north east lines,
    pattern color=cb_aquamarine,
    draw=cb_aquamarine]
    coordinates {(DDHselective,11.34) (DDHadaptive,14.76) (LWEselective,22.06) (LWEadaptive,32.23)};
\addplot+[style={cb_red,fill=cb_red!40}] coordinates {(DDHselective,48.31) (DDHadaptive,93.96) (LWEselective,2.45) (LWEadaptive,108.26)};

\legend{FLAD Training Time, Encryption Time, Aggregation Time}
\end{axis}
\end{tikzpicture} 
\end{adjustbox}
\caption{Time comparison of different phases within a round.}
\label{fig:timebench}
\vspace{-3mm}
\end{figure}



\section{Scheme generalizations}\label{sec:generalization}
Our scheme can be adapted to \ac{fl} schemes other than
 \ac{flad}; the same principles can be applied to enhance the security of a broader class of \ac{fl} algorithms, provided that the underlying scheme guarantees some properties required for compatibility with our cryptographic extension:
\begin{enumerate}[label=P\arabic*]
\item in the aggregation step, the mean, either weighted or arithmetic, is performed on each client's parameters;\label{property_aggregation}
\item all clients send a model to the server at every round;\label{property_all_clients}
\item the assignment of steps and epochs either does not depend on client secrets, but only on their aggregation; or it depends on secrets known only to the client running it, such as in~\Cref{alg: trainload}.\label{property_steps_and_epochs}
\end{enumerate}
\ac{flad} satisfies  (\ref{property_aggregation}) and 
(\ref{property_all_clients}). For other schemes,
(\ref{property_aggregation}) can be solved by adding a round of communication between each client and the server. In many FL schemes, e.g. FedAvg, the mean $w^t$ is performed on a different subset $C_t$ of the clients depending on the round $t$, that is
\begin{equation}
    w^t = \frac{\sum_{i\in C_t}k_iw^{t-1}_i}{\sum_{i\in C_t}k_i},
\end{equation}
where $k_i$ is the size of the $i$-th client's dataset.
This can be addressed in a naive way by changing the MIFE key in each round, exposing the scheme to \textit{decryption key mix-and-match attacks}. Instead, we can ask for each client $i$ to share with the server the MIFE encryption of the quantity
\begin{equation}
    \delta_i = \begin{cases}
        &k_i \ \ \text{if $i$ trained in the current round,}\\
        &0 \ \ \text{otherwise.}
    \end{cases}
\end{equation}
Note that the following holds:
\begin{equation}
    \frac{\sum_{i\in C_t}k_iw^{t-1}_i}{\sum_{i\in C_t}k_i} = \frac{\sum_{i\in C}\delta_iw^{t-1}_i}{\sum_{i\in C}\delta_i},
\end{equation}
where $C$ is the set of all the clients.
Thus, once the clients get the quantity $\sum_{i\in C}\delta_i$ through MIFE, they can send to the server the encryption of the quantity $\frac{\delta_i}{\sum_{i\in C}\delta_i}w^{t-1}_i$. Once the server decrypts this data with the usual decryption key associated with the vector $\mathbf{y} = (1,\dots,1)$, it finally gets the desired new model $w^t$.




Even if a \ac{fl} scheme does not satisfy (\ref{property_aggregation}, \ref{property_all_clients}, \ref{property_steps_and_epochs}), it is possible to slightly modify the scheme in such a way that 
\begin{itemize}
    \item this modification does not impact  performance, such as the number of rounds to achieve convergence or the accuracy of the final model, e.g. our modification in the epochs and steps assignment in FLAD (see \Cref{alg: trainload});
    \item the trade-off between security guarantees and performance loss satisfies the use-case.
\end{itemize}

\section{Conclusion and Future Work}\label{sec:conclusions}

In this paper, we have addressed security issues in the context of \ac{fl} applied to network security problems. 
Specifically, we have integrated \ac{mife} techniques with \ac{flad}, a recent adaptive \ac{fl} implementation for network intrusion detection. Moreover, we have discussed how our approach can be generalized to a broader class of FL schemes.

Despite preventing reconstruction attacks, our experimental results show that adding a layer of cryptography heavily impacts the memory and time performance of \ac{flad}, when security parameters are set to modern standards. However, these findings give a realistic benchmark for the performance of two standard choices such as DDH-based and LWE-based MIFE, and represent a major improvement compared to similar studies in the scientific literature.

Surprisingly, the selective version of LWE-based MIFE can outperform its DDH counterpart, while preserving moderate memory consumption and being post-quantum secure. Our work shows that adaptive LWE-based MIFE is costly in a realistic scenario. This encourages further study on different quantum resistant solutions, such as RLWE \cite{rlwe}. 

In a further study, we want to compare the performance of different post-quantum primitives and to test and propose new batch encryption  routines for client plaintexts. 

We also aim to investigate an extension of this approach to Federated Learning algorithms in which only clients in a random subset send their models in each round.

\bibliographystyle{IEEEtran}
\bibliography{bibliography}

\end{document}